\documentclass[ps,prd,preprintnumbers,superscriptaddress,nofootinbib,floatfix,twocolumn,notitlepage]{revtex4-2}
\pdfoutput=1 

\usepackage{dcolumn}

\usepackage{bm,amssymb,slashed,graphicx,multirow,soul,mathtools,xspace,array,tikz,amsmath,siunitx}
\usepackage[compat=1.1.0]{tikz-feynman}   
\usepackage{float}                   
\allowdisplaybreaks
\usepackage{ bbold }
\usepackage{caption,subcaption}
\usepackage{braket}
 \usepackage{hyperref}

\definecolor{nicered}{rgb}{0.7,0.1,0.1}
\definecolor{nicegreen}{rgb}{0.1,0.5,0.1}
\definecolor{violet}{rgb}{0.7,0.3,0.3}
\hypersetup{colorlinks,citecolor= nicegreen,linkcolor= nicered}

\newcommand{\lp}{\left(}
\newcommand{\rp}{\right)}

\newcommand{\beq}{\begin{equation} }
\newcommand{\eeq}{\end{equation}} 
\newcommand{\bi}{\begin{itemize} }
\newcommand{\ei}{\end{itemize} }

\definecolor{Red}{rgb}{1.,0.,0.}
\definecolor{Grn}{rgb}{0.,0.75,0.}
\definecolor{Blu}{rgb}{0.,0.,1.}
\definecolor{Pink}{rgb}{1,0.08,0.58}

\usepackage[T1]{fontenc} 

\usepackage{amsmath,amssymb,epsfig,ulem,color,slashed}
\allowdisplaybreaks  

\newcommand{\Xm}{X_{\rm max}}

\begin{document}

\title{\boldmath Resolution of (Heavy) Primaries in Ultra High Energy Cosmic Rays}

\author{Bla\v{z} Bortolato}
\email{blaz.bortolato@ijs.si}
\author{Jernej F. Kamenik}
\email{jernej.kamenik@cern.ch}
\affiliation{
 Jo\v{z}ef Stefan Institute, Jamova 39, 1000 Ljubljana, Slovenia \\
 Faculty of Mathematics and Physics, University of Ljubljana, Jadranska 19, 1000 Ljubljana, Slovenia
}

\author{Michele Tammaro}
\email{michele.tammaro@fi.infn.it}
\affiliation{INFN Sezione di Firenze, Via G. Sansone 1, I-50019 Sesto Fiorentino, Italy}

\date{\today}

\begin{abstract}
Measurements of Ultra-High Energy Cosmic Rays (UHECR) suggest a complex composition with significant contributions from heavy nuclei at the highest energies. We systematically explore how the selection and number of primary nuclei included in the analysis impact the inferred UHECR mass composition. Introducing a distance measure in the space of $\Xm$ distribution moments, we demonstrate that limiting the analysis to a few primaries can introduce significant biases, particularly as observational data improves. We provide lists of primaries approximately equidistant in the new measure, which guaranty unbiased results at given statistical confidence. Additionally, we explore consistent inclusion of nuclei heavier than iron and up to plutonium, deriving first observational upper bounds on their contributions to UHECR with the Pierre Auger Open Data.

\end{abstract}

\maketitle

%
\section{Introduction}
\label{sec:intro}
%

Measurements of Extensive Air Showers (EAS) at Pierre Auger Observatory (PAO)~\cite{ PierreAuger:2015eyc} and at Telescope Array~\cite{KAWAI2008221} indicate that the mass spectrum of Ultra-High Energy Cosmic Rays (UHECR) is best explained by a mixture of different primaries. Their inferred composition also depends on the cosmic ray energy: at lower energies ($E\sim10^{18}$ eV), it seems to be dominated by protons and possibly other light elements, while measurements at higher energies ($E\sim10^{19} - 10^{20}$ eV) point towards a significant fraction of heavy nuclei~\cite{PierreAuger:2014gko, Lipari:2020uca, Arsene:2020ago, PierreAuger:2017tlx}. 

The UHECR composition is closely related to other open questions in the field, regarding the sources of UHECRs, their acceleration mechanisms and their propagation in the galactic and intergalactic medium. 

Typically, the nuclei assumed to be part of UHECRs (apart from protons) span from stable isotopes of helium (He, $A=4$) to iron (Fe, $A=56$), as Fe has the largest binding energy per nucleon and it is likely the end product of stellar evolution. While heavier elements ($A>56$) are rarer in cosmic radiation, they can be produced at UHE, e.g. in GRB~\cite{Metzger:2011xs}, and have a similar propagation phenomenology as lighter elements~\cite{Epele:1998ia, Bertone:2002ks, Allard:2011aa, Zhang:2024sjp}. Thus it is reasonable to explore scenarios where these super heavy nuclei make up a fraction of the UHECR and to study their observational features at observatories like the PAO. Recent studies indicate the presence of $Z\sim 53$ primaries as a possible origin of the ``second knee'' in the UHECR energy spectrum~\cite{Gaisser:2013bla,Lv:2024wrs}.

At terrestrial observatories, the composition of UHECRs can be inferred by studying the longitudinal profile of an EAS, that is the intensity of fluorescent light emitted by nuclei in the atmosphere, typically nitrogen, excited by the passage of charged particles, and measured as a function of the s.c. slant depth ($X$) of the shower. In general, the longitudinal profile has a clear peak, at $\Xm$, corresponding to the point of maximum population of electrons and positrons in the shower evolution. Comparing observed data on $\Xm$ distributions at a particular energy against simulations for a set of primaries, one can infer the allowed fraction of each considered element in the observed collection of UHECRs. 

Past studies of UHECR composition have considered either a binned distribution of $\Xm$ or its global features such as its mean $\langle \Xm \rangle$ and width $\sigma({\Xm})$~\cite{PierreAuger:2014gko, Arsene:2020ago}. Typically the measured values of these were compared to simulations of a few chosen primaries. Recently~\cite{Bortolato:2022ocs}, we have have shown how a systematic description of the $\Xm$ distribution in terms of its central moments together with efficient inference techniques based on bootstrapping and nested sampling can be used to infer the full composition of all stable primaries up to iron. We have also shown that while the fractions of individual primaries in this case can be highly correlated and thus degenerate with each other, one can still derive robust confidence intervals for fractions of all primaries {\it heavier than} any chosen nuclear mass or atomic number.

In this work we extend our previous studies of UHECR composition by addressing two main questions concerning the choice of considered primaries when inferring composition, namely how does the (1) range of primaries and (2) their total number affect the inferred composition confidence intervals. This is particularly important for both present and future studies. Firstly, the current results are based on the (strong) assumption that a mixture of only four primaries, typically (p,He,C/N/O,Fe)~\cite{PierreAuger:2023bfx}, describes well the UHECR mass spectrum. Secondly, it is warranted in order to to derive bounds on possible fractions of primaries beyond iron, because a brute-force approach of including all stable enough primaries up to uranium ($Z=92$) or plutonium ($Z=94$) is computationally extremely expensive using existing techniques. To this end we first introduce a natural metric in the space of $\Xm$ distribution moments. Because simulated $\Xm$ distributions of primaries have intrinsic uncertainties, this allows on one hand to identify a-priori practically (given finite event statistics) indistinguishable primaries, and on the other allows us to include only distinguishable primaries in the inference procedure. Conversely, we study in detail using mock data, how omission of distinguishable primaries can potentially lead to biased inference of primary UHECR composition. In particular, we show that mass composition results of existing studies using the full PAO statistics but with limited number of primaries below $Z=26$ are potentially biased. Finally we explore the effects of including primaries beyond $Z=26$ in the composition and in particular derive first upper bounds on fractions of UHECR primaries up to plutonium using the PAO open dataset.

The remainder of the paper is structured as follows: in Sec.~\ref{sec:methods} we introduce our methodology and data, in particular the PAO open dataset, the CORSIKA EAS simulation framework, the relevant observables -- moments of $\Xm$ distributions, and our statistical inference procedure.  In Sec.~\ref{sec:compositionVSdistance} we study the implications of inferring composition using different numbers of primaries. We first define a distance measure in the space of $\Xm$ distribution moments and show how it can be used to define distinguishable primaries. We apply this measure to select primaries to be included in the inferred mixture. We test this procedure both using mock data as well as the PAO open dataset, where we also infer composition including primaries beyond iron and up to plutonium. Our main conclusions are summarised in Sec.~\ref{sec:conclusions}, while the Appendices~\ref{app:Interpolation} and~\ref{app:OtherBin} contain some auxiliary computations, as well as additional tables and plots, respectively.  

%
\section{Methodology}
\label{sec:methods}
%

\paragraph{Open Data}: we use publicly available data from the 2021 Pierre Auger Open Data (PAOD) release~\cite{the_pierre_auger_collaboration_2021_4487613}. This dataset consists of 10\% of the full recorded data by the PAO. In particular, we focus on so called hybrid showers: events recorded by both the Surface Detectors (SD) and the Fluorescent Detectors (FD). There are 3348 hybrid showers, distributed in the reconstructed primary energy as $\sim E^{-2.6}$, from $E_{\rm min} = 0.65$ EeV to $E_{\rm max} = 63$ EeV. We select the energy bin $E\in[0.65, 1]$ EeV, containing 934 events, and use it in all our remaining discussion. In this way we work with a fairly large statistical sample within a small energy bin. For completeness, we include the respective results in the energy bin $E\in[1,2]$ EeV, encompassing 1234 events, in Appendix~\ref{app:OtherBin}.

\paragraph{Simulations}: we simulate EAS with CORSIKA v7.7550~\cite{1998cmcc.book.....H}. To achieve fast and reliable simulations of longitudinal shower profiles, we employ the CONEX option. In conjunction with the EPOS hadron interaction model, this option allows to include primary nuclei up to $A=295$. Note however that composition results can strongly depend on the choice of the hadronic model~\cite{Bortolato:2022ocs}. Thus, our results cannot explore the potential effects of different parametrizations of UHE hadronic cross sections.
We simulate $10^4$ showers per primary, with the energy uniformly distributed in the chosen energy bin. The atomic number $A$ of the primary is taken as the one of the most abundant stable isotope; we thus unambiguously indicate each primary nucleus by its atomic number $Z$. We simulate all primaries from $Z=1$ (proton) to $Z=26$ (iron); in addition, we simulate showers for heavier primaries with $Z=$ 27, 28, 29, 30, 34, 39, 40, 42, 43, 44, 47, 50, 57, 58, 64, 72, 81, 82, 91, 92, 94. We chose to simulate a set of primaries with atomic number immediately beyond iron, and additionally a set of primaries that cover the interval up to Plutonium, $Z=94$, which is the heaviest that can be simulated with CONEX. 
Finally, we extract the value of $\Xm$ from each simulated shower by fitting the longitudinal profile to a Gaisser-Hillas curve~\cite{1977ICRC....8..353G}. The set of $\Xm$ per primary furnishes a distribution $p(\Xm|Z)$.

\paragraph{Observables}: we describe the resulting $\Xm$ distribution by its moments, defined as
\beq\label{eq:moments}
\langle\Xm^n\rangle\ = \frac{\int p(\Xm|Z) \Xm^n{\rm d}\Xm}{\int p(\Xm|Z) {\rm d}\Xm}\,.
\eeq
From this expression we can obtain the central moments as
\beq
\begin{split}
z_n &= \langle (\Xm - \langle \Xm \rangle)^n \rangle \\
&= \sum^n_{k = 0}\binom{n}{k} \langle \Xm^{n-k} \rangle  (-1)^{k} \langle \Xm \rangle^k\,.
\end{split}
\eeq
The number $n$ of moments needed to infer the composition depends on the interplay of statistical and systematic uncertainties, which we discuss in the next paragraph. We have shown in Ref.~\cite{Bortolato:2022ocs} that with the data available in the Auger Open Data, $n=3$ is sufficient and additional moments do not improve the results.

\paragraph{Uncertainties}: We consider both systematic and statistical uncertainties on $\Xm$. The former originate from the finite resolution of the fluorescent detector~\cite{PierreAuger:2014sui}; we include these by reweighting the distribution $p(\Xm|Z)$ by the detector smearing~\cite{Bortolato:2022ocs}. The latter are a consequence of the limited statistics available, either in the simulation or in the observed data. We take these into account by means of bootstrapping~\cite{EfroTibs93}, that is repeatedly sampling sets of $\Xm$ from $p(\Xm|Z)$ and computing moments of these sets; the spread of the moment distributions obtained will depend on the total number of events available. 

The next observational run of Auger Prime~\cite{Stasielak:2021hjm} is expected to deliver a larger and more precise dataset. In the following we perform projections to larger statistics by assuming that the central values of the measured $\Xm$ moments remain unchanged, while the variances of moment distributions scale with the \textit{statistical multiplier} $f$ as $1/f$, when increasing the total available statistics by a factor of $f$.\footnote{An example of how to effectively enhance existing $\Xm$ statistics by exploiting correlations with SD data has been shown in Ref.~\cite{Bortolato:2023wsk}, where a more general definition of $f$ is also provided.} While such scaling is formally only valid for statistical uncertainties, at the current and even projected PAO statistics, also the dominant systematic uncertainties are expected to effectively scale with statistics. Namely, the detector resolution is calibrated using (calibration) data and the resulting uncertainties are statistics dominated~\cite{PierreAuger:2014sui}.

\paragraph{Inference procedure}: we follow the steps of our work, Ref.~\cite{Bortolato:2022ocs}, to infer the composition of UHECR, $w$. For a given mixture of primaries, the log-likelihood is
\beq\label{eq:likelihood}
\ln{\cal L}(w) = \int \ln\Big[ p\lp z| \mu_w, \Sigma_w \rp \Big]~ p\lp z|\tilde\mu,\tilde\Sigma \rp ~{\rm d}z\,,
\eeq
where $p\lp z|\tilde\mu,\tilde\Sigma \rp$ is the moment distribution of the Auger data, while $p\lp z| \mu_w, \Sigma_w \rp$ is the distribution of moments for simulated momenta, weighted by the composition $w$. 
The number of free parameters depends on the set of $Z$ allowed to be in the mixture, which is worked out in Sections~\ref{sec:Distance} and ~\ref{sec:compositionVSdistance}. The best fit composition can be obtained by maximizing the log-likelihood in Eq.~\eqref{eq:likelihood}. However, the typical number of free parameters and thus the dimensionality of the likelihood manifold is large. In order to be able to effectively extract the confidence intervals of our best fits, that is the shape of $\ln{\cal L}$ around the maximum, we use Nested Sampling (NS) techniques~\cite{doi:10.1063/1.1835238, UNgithub}. The latter allow us to sample high-dimensional likelihoods, with up to 100 free parameters, thanks to a large reduction in computational time with respect to traditional sampling methods.

We present the results as a \textit{cumulative composition}: for each $Z$ included in the fit, we show the allowed fraction of primaries heavier than $Z$. This method yields a reliable presentation of the results, including well defined confidence intervals that account for correlations among individual primaries' fractions. It also allows to compare composition results with different numbers of primaries since, contrary to individual primary fractions, the cumulative fraction values for any $Z$ are always well defined and have the same statistical meaning irrespective of the actual primaries included in the fit (as long as protons are included).

%
\section{Composition with different numbers of primaries}
\label{sec:compositionVSdistance}
%

The procedure of inferring the composition of UHECR, introduced in Ref.~\cite{Bortolato:2022ocs} and summarized in Sec.~\ref{sec:methods} raises an important question: what is the minimal number of composition components required to obtain an unbiased fit to data? Namely, defining the composition vector $w=(w_{Z_1},w_{Z_2},\dots,w_{Z_k})$, we wish to quantify the minimal length $k$ and find a suitable list of $Z_k$ primaries, such that the inference procedure yields (cumulative) composition fractions consistent with their truth level values. The set of primaries (p,He,C/N/O,Fe), that is $N_L=4$, has been largely assumed in the literature~\cite{Lipari:2020uca,PierreAuger:2023bfx} to represent the UHECR mass spectrum. Previous works, see Refs.~\cite{Arsene:2020ago,Arsene:2021inm}, have however shown that the goodness of fit for the UHECR composition is strongly affected by the number of primaries chosen, and that it generally improves when including more intermediate elements, up to $N_L = 8$. Here we extend these studies in a systematic way, employing a distance measure in the space of $\Xm$ moments to provide a quantitative criterion for the selection of primaries.

%
\subsection{Primary Distance in Space of $\Xm$ Moments}
\label{sec:Distance}
%
The decomposition of the $\Xm$ distributions in terms of a series of central moments allows us to quantify the separation between different primaries $Z$.
Given two primaries $Z_1$ and $Z_2$, we define the distance 
\beq\label{eq:Distance}
d_n^2(Z_1,Z_2) \equiv (\mu_{Z_1} - \mu_{Z_2})^T (\Sigma_{Z_1} + \Sigma_{Z_2})^{-1} (\mu_{Z_1} - \mu_{Z_2}) ,
\eeq
where $n$ indicates the number of moments used. Here $\mu_{Z_1}, \mu_{Z_2}$ are the $n$-vectors of moment means and $\Sigma_{Z_1}, \Sigma_{Z_2}$ the respective $n\times n$ covariance matrices\footnote{Notice that this definition respects the properties of a distance, thus it is equivalent to other more familiar distance definitions, e.g. the Hellinger distance.}. This definition builds upon the property of the bootstrapped distributions of moments, that they quickly approach normal distributions. The same feature allows for a straightforward statistical interpretation of $d_n^2$: the distance behaves as a chi-squared distribution with $n$ degrees of freedom, $d_n^2\sim\chi_n^2$, thus its value can be converted to a confidence level of distinguishing among two sets of moments. For example, if we fix $n=1$ moments and the distance between two primaries is $d_1^2 \sim \chi_1^2 \simeq 2.7$, then we can interpret this as $90\%$ confidence that the compositions consisting of the two primaries are in principle distinguishable using measurements of only the first moment. 

Since $\Xm$ distributions of individual primaries cannot be inferred directly from data, distances $d_n^2$ need to be estimated using simulations. The main systematics in calculating the moments and in particular their means is then expected to come from the parametrization of UHE hadronic interactions, i.e. the choice of the hadronic model. In addition however, the corresponding covariance matrices also scale with the simulation statistics. This means that the separability of pairs of primaries crucially depends on the assumed (simulated) data statistics.  In particular, for $N$ simulated or measured events, the distances $d_n$ scale as $\sqrt{N}$. The scaling remains true even in the presence of systematic uncertainties on the $\Xm$ measurements. As detailed in Ref.~\cite{PierreAuger:2014sui} and discussed in Sec.~\ref{sec:methods}, the main systematics namely come from the limited statistics of the observed data. 

\begin{figure}[t!]
    \centering
    \includegraphics[scale = 0.8]{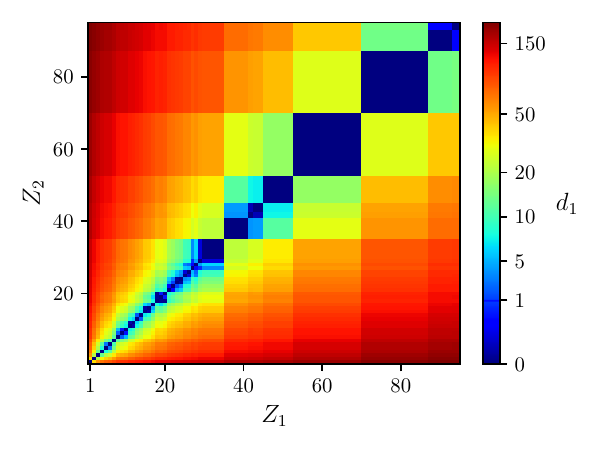}
    \includegraphics[scale = 0.8]{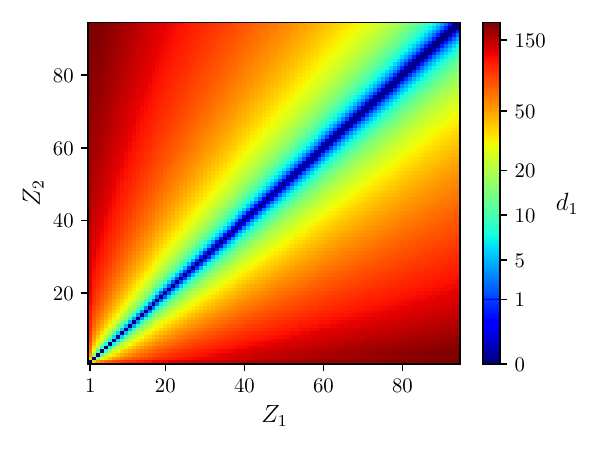}
    \caption{{\bf Top:} distance $d_1$ for each pair of simulated primaries. {\bf Bottom:} distance $d_1$ for each pair of primaries, using the interpolating functions in Eq.~\eqref{eq:interpolation_functions}. 
    Note that the scale is linear for $d_1 \in [0,1]$ while it is logarithmic above this interval.
    See text for details. 
    }
    \label{fig:distance_matrix_bin1}
\end{figure}

In the top plot of Fig.~\ref{fig:distance_matrix_bin1} we show the $n=1$ distances $d_1$ for the full list of our simulated primaries in the chosen energy bin, assuming the same statistics as PAOD. As expected, the distance for a fixed $Z_1$ grows with the difference $|Z_2 - Z_1|$ as the distributions (and consequently their moments) of $\Xm$ drift apart for more distant primaries. Somewhat surprisingly however, we find that the distance $d_n$ between any two primaries is to a very good approximation independent of the number of moments considered $n$. This observation can be traced back to the covariance matrices $\Sigma_Z$, which exhibit a high degree of correlation between the different moments, as first observed already in Ref.~\cite{Bortolato:2022ocs}.  
This property allows us to disregard higher moments in the context of distances, and consider only the first moment $z_1$ for all the simulated primaries. In addition, the monotonic nature of $z_1$ allows us to employ an analytic interpolation formula for the mean and variance of $z_1$ as functions of $Z$ (see Appendix~\ref{app:Interpolation} for details) in order to compute $d_1$ for all pairs of primaries (including those not explicitly simulated) up to $Z=94$, as shown in the bottom plot of Fig.~\ref{fig:distance_matrix_bin1}.

Leveraging on the statistical interpretation of the distance between primaries and making use of these interpolating functions, we can obtain lists of approximately \textit{equidistant} primaries. We fix the first element to always be proton, $Z=1$, and choose a step distance $d_{0}$. We then iteratively find the next element of the list as the one whose distance from all previous is at least $d_1 \geq d_0$. Each choice of $d_0$ thus directly corresponds to a set of $Z$ and defines the level of separation between neighbouring primaries in the shower composition. 

\begin{figure}[t]
    \centering
    \includegraphics[scale = 0.8]{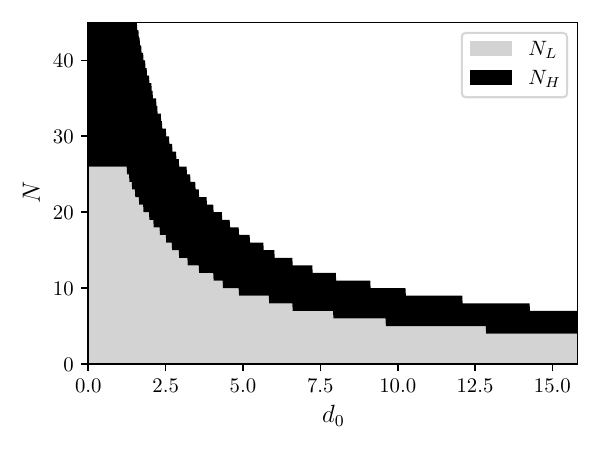}
    \caption{Stacked $N_L$ (gray) and $N_H$ (black) values as a functions of the distance threshold $d_0$, assuming $10^3$ events per primary in the energy bin $[0.65,1]$ EeV. 
    }
    \label{fig:NvsD}
\end{figure}

We show some examples in Table~\ref{tab:list_primaries_lower_bin1}, assuming $10^3$ events per primary, where we also indicate the number of primaries with $Z\leq26$ and $Z>26$ as $N_L$ and $N_H$ respectively. Note that a set with $N_L=4$, as widely employed in the literature, would correspond to a relatively large distance step, $d_0=16.6$ at these statistics, while asking for at most $d_0<2$ separation leads to $N_L=20$. Due to the scaling of $d_n$ with assumed primary statistics, these lists can be used at any given statistics if we also rescale the threshold value $d_0$ by the corresponding statistical multiplier $\sqrt{f}$. For example, by increasing the dataset tenfold, $f=10$, to $10^4$ events per primary, the same lists are obtained by imposing the new thresholds $d_0' =\sqrt{f}d_0\sim3d_0$. 
For completeness, we show the behaviour of $N_L$ and $N_H$ as functions of $d_0$ assuming $10^3$ events per primary in Fig.~\ref{fig:NvsD}. 

We note that the distributions of $\Xm$ moments change with CR energy and thus both $N_{L,H}$ as well as the lists of $d_0$-equidistant primaries can change for the same $d_0$ in different energy bins, see e.g. Table~\ref{tab:list_primaries_bin2}, so these should be recomputed for any specific CR energy range considered.

Finally, in any measured CR event sample of unknown composition, only the total number of registered showers is known a-priori, and not the statistics of individual primary components. Thus, in practice one can only compute an upper bound on $d^2_n(Z_1,Z_2)$ by assuming that the dataset set is composed of only two primaries ($Z_1$ and $Z_2$) of equal fractions. The actual separability within the dataset will then always be lower and depend on the (unknown) true composition. Our $d_0$ values and lists in Tables~\ref{tab:list_primaries_lower_bin1} and~\ref{tab:list_primaries_bin2} can thus be considered as very conservative if applied to event samples of given total statistics. They nonetheless provide meaningful benchmarks, as we demonstrate using simulated mock data examples of known composition in the next section.

\begin{table}[t!]
    \centering
    \begin{tabular}{c|c|p{60mm}}
        $d_0$ & $N_{L}$ ($N_{H}$) &  List of atomic numbers $Z$ \\ \hline
        16.6 & 4 (2) &  1,  3, 10, 24, 52, 94    \\ \hline
        6.4 & 8 (5) &  1,  2,  4,  6,  9, 13, 19, 27, 37, 50, 67, 89  \\ \hline
        4.0 & 12 (7)&  1,  2,  3,  4,  5,  7,  9, 11, 14, 17, 21, 26, 31, 37, 44, 53, 63,
       75, 89 \\ \hline
        2.8 & 16 (10) &  1,  2,  3,  4,  5,  6,  7,  8,  9, 11, 13, 15, 17, 20, 23, 26, 30,
       34, 39, 44, 50, 57, 64, 72, 81, 91 \\ \hline
        2.0 & 20 (14) & 1,  2,  3,  4,  5,  6,  7,  8,  9, 10, 11, 12, 13, 14, 16, 18, 20,
       22, 24, 26, 29, 32, 35, 38, 42, 46, 50, 54, 59, 64, 70, 76, 82, 89   \\ \hline
        1.3 & 24 (21) & 1,  2,  3,  4,  5,  6,  7,  8,  9, 10, 11, 12, 13, 14, 15, 16, 17,
       18, 19, 20, 21, 22, 24, 26, 28, 30, 32, 34, 36, 38, 40, 43, 46, 49,
       52, 55, 58, 62, 66, 70, 74, 78, 83, 88, 93
    \end{tabular}
    \caption{Lists of approximately \textit{equidistant} primaries in the chosen energy bin. The first column shows the chosen step $d_0$ assuming $10^3$ events per primary, while the second indicates the values of $N_L$ and $N_H$ obtained. The last column lists all selected $Z$.
    }
    \label{tab:list_primaries_lower_bin1}
\end{table}

%
\subsection{Composition Inference on Mock Data}
\label{sec:compositionVSdistance:fake}
%
We demonstrate the implications of different $d_0$ based $w$ length choices for the mixture fit results using specific mock compositions of UHECR based on simulated data. We first select a set of primaries and their respective fractions to represent the truth-level composition, $w_{\rm true}$.
We compute the moments of the truth-level $\Xm$ distribution by randomly sampling 1000 events, apportioned following $w_{\rm true}$. In order to ensure consistent truth level compositions at large values of $f$, we do not separate these events from the rest of the simulations. Otherwise, the computed values of the distribution moments would be slightly different from the ones computed using all events; thus at large $f$, this difference could lead to significant deviations from $w_{\rm true}$.

We study two example compositions:
\begin{enumerate}
    \item[Ex1] $w_1 = w_{10} = 1/2$, that is we assume the mass spectrum consists of equal fractions of only two primaries, $Z=1$ and $Z=10$;
    \item[Ex2] $w_1 = w_{2} = \dots = w_{10} = 1/10$, that is we take equal fractions of the lightest ten primaries.
\end{enumerate}
For each example, we infer the composition using the lists of light primaries\footnote{We discuss the effect of introducing additional $N_H$ heavy primaries beyond $Z=26$ in Section~\ref{sec:composition:Heavy}.} with $N_L = 4,8,12,16,20,24,26$, as shown in Table~\ref{tab:list_primaries_lower_bin1}, and for several values of the statistical multiplier $f$. The results are presented in Fig.~\ref{fig:fakecomp:logL} for Ex1 (top) and Ex2 (bottom), respectively. On the left side we show the truth level cumulative composition $w_{\rm true}$ (in black lines),\footnote{We have checked explicitly that the best fit composition for $N_L=26$ in both examples is consistent with the true composition.} and compare it to the 95\% confidence intervals for three choices of $N_L$ (note that we only show the cumulative fractions at the $Z$ values included in $w$ for each choice); on the right side we plot the differences of maximum $\ln{\cal L}$ values (corresponding to best fit compositions) with respect to the value obtained with $N_L=26$, $\ln{\cal L}_{26}$, that is by considering all light primaries in the composition.

{In both examples and for all choices of $N_L$, cumulative composition confidence intervals shrink with increasing statistics ($f$), as expected. However, for too small $N_L$  (depending on composition and statistics)
the $95\%$ confidence intervals fail to cover the truth-level values $w_{\rm true}$ thus indicating biased inferred compositions.

An extreme example is given by the $N_L=4$ case in Ex2. In the left plot, the red bars show an increasing precision on the cumulative fractions with increasing statistics. However, the intervals are systematically pulled away from the truth-level values indicated by the black line. A significant deviation is present already at $f=1$, that is assuming $10^3$ measured events in total, and it quickly grows with larger statistics.   The $N_L=4$ case thus leads to an underestimation of the heavier primary fractions and to biased results in general, even for event statistics comparable to PAOD. At the same time the relative goodness of fit degrades significantly with increasing $f$, as seen in the right plot.

\begin{widetext}

\begin{figure}[!h]
    \centering
    \includegraphics[scale = 0.7]{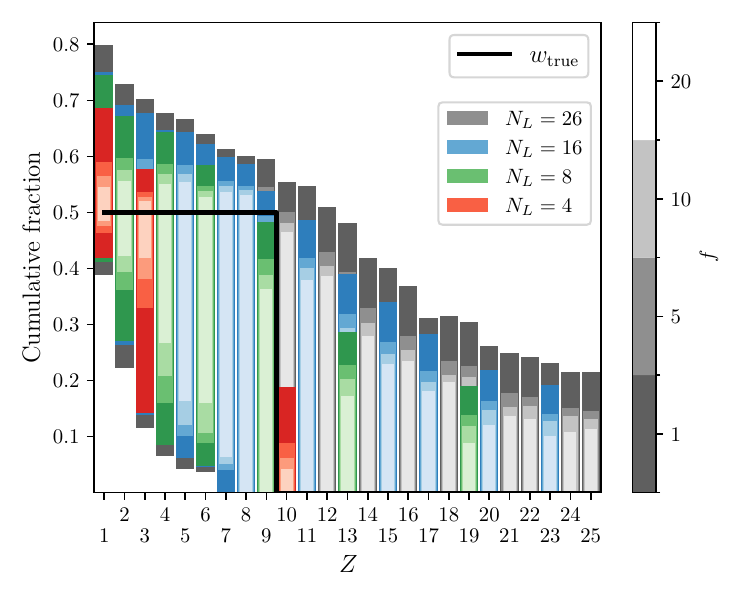}
    \includegraphics[scale = 0.8]{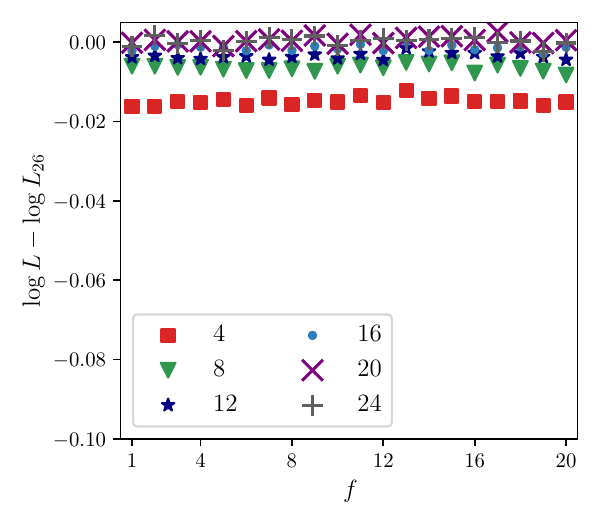} \\
    \includegraphics[scale = 0.7]{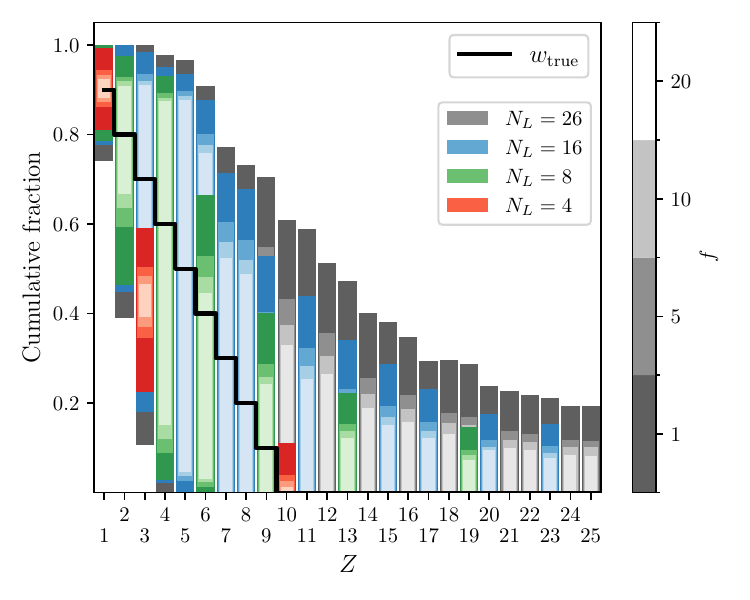}
    \includegraphics[scale = 0.8]{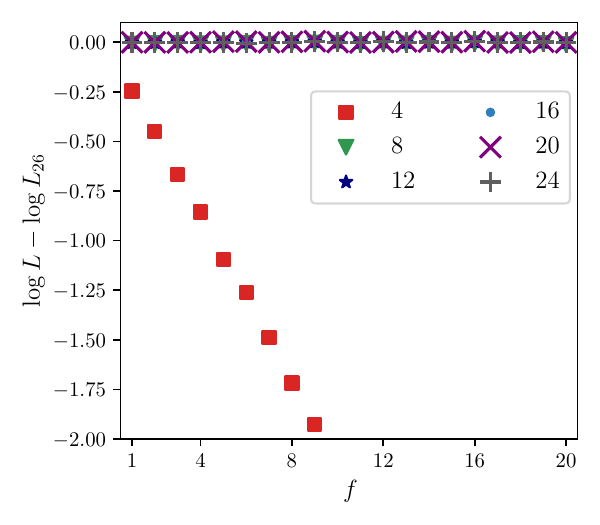}
    \caption{Cumulative composition confidence intervals and relative goodness of fit for Ex1 ({\bf top}) and Ex2 ({\bf bottom}) mock UHECR compositions. {\bf Left}: the black line indicates the truth-level composition, while colored bars indicate 95\% confidence level intervals for the respective primaries. Red, green, blue and gray bars refer to the inference with $N_L=4,8,16,26$ primaries respectively, while shades of the same color indicate the multiplier $f$, as specified by the legend on the right. {\bf  Right}: Difference of $\ln{\cal L}$ values of best fit compositions, with $N_L =  4,8,12,16,20,24$, with respect to the case $N_L=26$. See text for details. 
    }
    
    \label{fig:fakecomp:logL}
\end{figure}

\end{widetext}

On the other hand, we observe that this behavior is not universal and that the relative goodness of fit does not necessarily degrade significantly with increasing $f$. It can also remain comparable to the full $N_L=26$ case even as the composition results become biased. This can be most clearly seen for the  $N_L=8$ case in Ex1, where the log likelihood difference remains almost constant even as the fraction of the heavier element in the composition is significantly underestimated. Interestingly, in Ex1, the same happens even for $N_L = 16$, where at $f=5$ the fit starts to underestimate the heavier primary fraction and thus actually yields biased composition results. Note that this choice of $N_L$ corresponds to $d_0\simeq 6$ assuming $5\times 10^3$ events per primary. Although the true composition in Ex1 might be considered extreme, it nonetheless demonstrates that existing mass composition studies based on the full PAO dataset which all consider $N_L<16$, potentially yield  $95\%$ confidence intervals of primary fractions which do not cover their true values. }

%
\subsection{Application to Pierre Auger Open Data}
\label{sec:compositionVSdistance:real}
%

\begin{figure}[t!]
    \centering
    \includegraphics[scale =0.7]{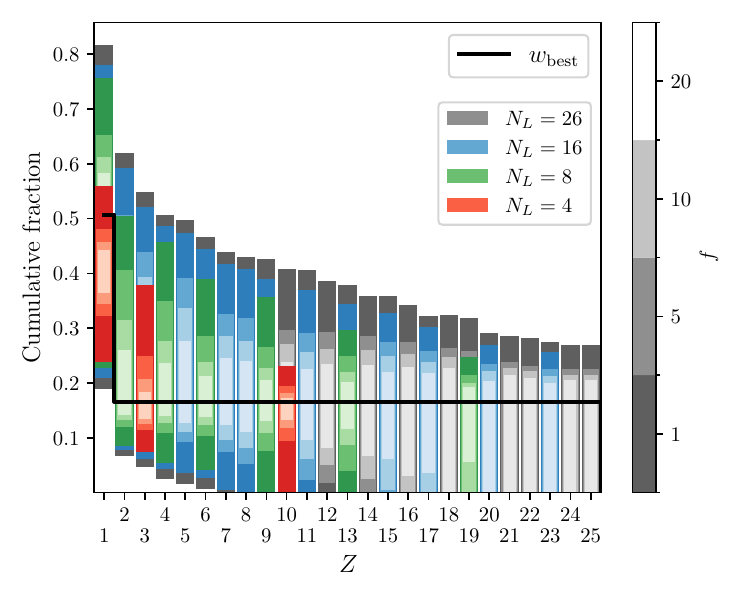}
    \caption{Cumulative fraction of primaries, with $95\%$ CL, obtained with $N_L=4,8,16,26$ (red, green, blue, gray), using PAOD. The solid black line indicates the best fit obtained with $N_L=26$.}
    \label{fig:composition:Real:onlyLight}
\end{figure}

Building upon the mock data examples studied in the previous Section, we next infer the composition of UHECR from PAOD in the chosen energy bin. We show our results for the cumulative fractions in Fig.~\ref{fig:composition:Real:onlyLight}, for different choices of $N_L$ and at various projected statistics factors $f$ (actual PAOD statistics correspond to $f=1$). Note again that for the moment we are restricting the procedure to lists of primaries up to iron, $Z\leq 26$ (and so $N_H=0$). Contrary to mock simulated data, it is of course impossible to know a priori the true mass composition of the UHECR spectrum. However, we can use the results of the full $N_L=26$ fit, in particular its maximum likelihood composition $w_{\rm best}$, as reference against which we compare other $N_L$ choices. We observe that a familiar pattern emerges: including only $N_L=4$ primaries, the fit tends to underestimate the cumulative fractions obtained with the full 26 primaries, especially the total fraction of nuclei with $Z>1$. Even more pronounced however is the underestimation of the $95\%$ confidence intervals, very similar to what is observed in mock data, where it signals a biased composition.
The $N_L = 16$ fit results instead are already well compatible with the reference $N_L = 26$ case including the total coverage of the $95\%$ CL intervals, at least at PAOD statistics (for $f=1$). 

%
\subsection{Bounds on Admixtures of Nuclei Beyond Iron}
\label{sec:composition:Heavy}
%

Finally, we study the effect of including possible ``heavy'' primaries, that is nuclei with $Z>26$. We have seen that, in order to have an unbiased estimation of the cumulative fraction of primaries in PAOD, we require at least $N_L=16$. We can then use the same procedure to extend our lists to include approximately $d_0$-equidistant primaries up to $Z \leq 94$, as shown in Table~\ref{tab:list_primaries_lower_bin1}.

\begin{widetext}
    
\begin{figure}[t]
    \centering
    \includegraphics[scale = 0.8]{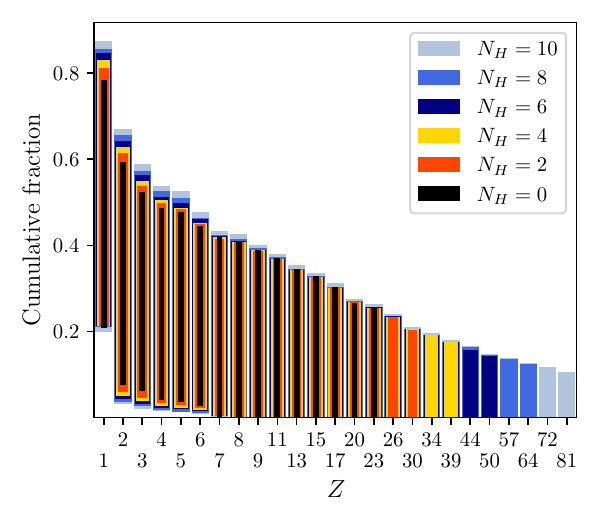}\hspace{1cm}
    \includegraphics[scale = 0.8]{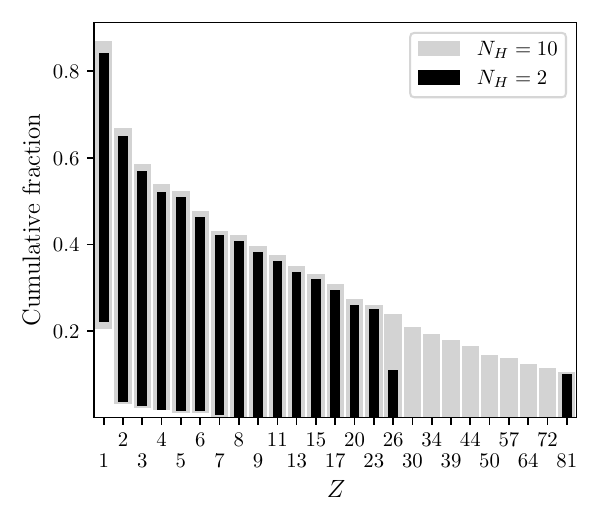}
    \caption{Cumulative fraction of primaries in PAOD at $95\%$ CL.  {\bf Right}: Results with $N_L=16$ (black), and adding $N_H = 2,4,6,8,10$ (red,yellow,dark blue, light blue, cyan) sequential $d_0$-equidistant heavy primaries. {\bf Left}: Results with $N_L=16$ and  $N_H=2$ ultra-heavy primaries ($Z=81$ and $Z=91$) (black), and $N_H=10$ $d_0$-equidistant  heavy primaries (gray). See text for details.}
    \label{fig:composition_heavy}
\end{figure}

\end{widetext}

To portray the effect of adding $N_H$ heavy primaries on the likelihood, we first infer the composition after adding two consecutive $d_0$-equidistant heavy primaries at a time, following the list for $N_L=16$ in Table~\ref{tab:list_primaries_lower_bin1}; the results are shown in the left-hand plot of Fig.~\ref{fig:composition_heavy}. The bounds on cumulative fractions do not change significantly, as it can be seen by comparing the black bands ($N_H = 0$) to the colored ones; the inferred CL intervals increase only slightly, due to the addition of new free parameters with the same statistical sample size. 
{As a direct result, we find that the fraction of primaries with $Z > 26$ in the energy interval $E\in [0.65,1]$~EeV is within $95\%$ CL bounded by
\beq
w(Z > 26, E\in [0.65,1]~{\rm EeV}) \leq 24\%,
\eeq
while for $E\in [1, 2]$~EeV we get
\beq
w(Z > 26, E\in [1,2]~{\rm EeV}) \leq 18\%.
\eeq
By construction, the upper bounds on heavy primaries are decreasing monotonically with increasing $Z$. Consequently, bounds on so-called super heavy nuclei, such as uranium and plutonium can be much tighter than the above values. As an illustration, we thus also make a conservative estimation on fractions of such primaries as
\begin{align}
w(Z > 81, E\in [0.65,1]~{\rm EeV}) &\leq 10\%, \nonumber \\
w(Z > 85, E\in [1,2]~{\rm EeV}) &\leq 6\%,    
\end{align}
both at $95\%$ CL. 
These values are obtained using the full $d_0$-equidistant range of $N_L=16$ and $N_H=10$ primaries on PAOD (see Tables~\ref{tab:list_primaries_lower_bin1} and~\ref{tab:list_primaries_bin2} for the lists).
}

A different outcome would be obtained if instead we only added super heavy primaries, e.g. $Z= 81, 91$, to the original 16 primaries with $Z \leq 26$. This is shown in the right-hand plot of Fig.~\ref{fig:composition_heavy}. In particular, we observe a substantial underestimation of the iron fraction compared to the results with $d_0$-equidistant primaries: both for $N_H=10$, as well as with $N_H=2$, where the consecutive heavy primaries $Z=30$ and $Z=34$ are included in the fit. We thus conclude that at least when applied to PAOD, the method of including consecutive $d_0$-equidistant primaries with $Z>26$ yields reliable upper bounds on cumulative fractions of heavy elements in UHECRs even when the full range of heavy nuclei is not included in the fit. Conversely, including only ultra-heavy primaries in the fit in addition to $Z\leq 26$ can lead to biased composition results.

%
\section{Conclusions}
\label{sec:conclusions}
%

In studies of the UHECR mass spectrum, the vector of primary composition fractions, $w$, fundamentally drives the inference procedure: its size defines the dimensionality of the problem, and as a consequence the computational power required to address it; the specific primaries that form $w$ potentially impose a strong prior on the inferred composition, which in turn can lead to biases in the fit results. It is thus important for the experimental collaborations determining the mass spectrum to base their procedures upon a solid and quantitative framework for selecting optimal $w$.

In order to provide such framework, we can leverage the discriminating power of the $\Xm$ distributions, to define a measure of distance between primaries in the space of $\Xm$ moments. By construction, the distance gives the statistical significance to distinguish between two primaries in a composition, as function of the statistical sample size. In other words, for normally distributed moments, this distance is equivalent to the significance, in units of $\sigma$, that the two sets of moments are distinct from each other. Furthermore, at least for the set of simulated primaries, this distance is well approximated by including only one (i.e. the first) moment. Consequently, we can build lists of approximately \textit{equidistant} primaries for the vector $w$; for convenience, we separate these primaries in ``light'', with $Z\leq26$, and ``heavy'', with $Z>26$, and indicate their numbers in the lists as $N_L$ and $N_H$. Taking $10^3$ events per primary as an example relevant for the statistics available in the PAOD, a minimum distance of $d_0\sim 3$ between primaries implies $w$ to be at least of length $N_L = 16$. Given the typical scaling with the statistical sample size, requiring the same $d_0$ with the full PA statistics would lead to even larger dimensionality for $w$.

We first tested this framework on two sets of simulated mock compositions. By fixing the underlying true $w$, we can observe the effect of imposing different $N_L$ on the inferred composition. The examples demonstrate that too low values of $N_L$ for given sample size (or equivalently too high $d_0$) in general lead to biased composition fraction estimates with underestimated uncertainties, see Fig.~\ref{fig:fakecomp:logL}. {As noted previously~\cite{Arsene:2020ago,Arsene:2021inm}, the relative goodness of fit, computed as the maximum value of the log-likelihood, in some cases tracks this increasing discrepancy of low $N_L$ results with respect to sufficiently large $N_L$ values. However we find that this behavior is not universal and that in general a high goodness of fit score does not guarantee unbiased composition results.}

A similar pattern is observed when applying the same strategy to PAOD, shown in Fig.~\ref{fig:composition:Real:onlyLight}; here the $N_L=4$ best fits seem to underestimate the fraction of primaries other than protons, with respect to the $N_L=16$ or $N_L=26$ results. The effect is expected to be statistically significant for the existing full PA dataset. {We thus recommend a minimum of $N_L=16$ $w$ components for the inference of the UHECR mass composition}, assuming that no elements heavier than iron are present.

By lifting the assumption that $Z\leq26$, we can repeat the procedure including so-called ``heavy'' primaries. We explored possible ways to add $N_H$ heavy primaries to $w$. The results shown in Fig.~\ref{fig:composition_heavy} indicate that adding very distant primaries; in the example presented, the introduction of only $Z=91$ and $Z=94$ in addition to $Z\leq 26$ strongly biases the fit. Fortunately, the bias can be reliably avoided by instead including sequential \textit{equidistant} heavier primaries starting from a given $N_L$ set.

At cosmic ray energies considered in this work, no heavy elements are actually expected, as the data favors a mixture of protons and lighter nuclei. Using PAOD we nonetheless put first observational bounds on the presence of primaries heavier than iron, as well as more specifically on elements heavier than uranium and plutonium in UHECRs in the energy range $E\in [0.65,2]$~EeV. The same analysis cannot be carried out at higher energies, as the PAOD does not contain a meaningful number of events above $\sim2-5$ EeV. 

Our work could be extended in several directions. Besides applying our methodology to the full PA (and Telescope Array) datasets, the determination of an optimal composition vector $w$ for a given CR dataset of unknown composition remains an open problem. We have argued that, while necessarily conservative, interpreting $d_0$ values computed with individual primary statistics comparable to the full dataset as resolution significances (measured in normal distribution sigmas) gives meaningful benchmark values to determine both the size and the relevant components of $w$. However, the potential reduction of computational cost as well as inferred composition uncertainties due to a lower number of free parameters when working with a shorter composition vector $w$ motivate further optimization, which we leave for future work. 

\section*{Acknowledgments}
B.B. and J.F.K. acknowledge the financial support from the Slovenian Research Agency (grant No. J1-3013 and research core funding No. P1-0035).

\bibliographystyle{h-physrev}
\bibliography{references}

\newpage
\clearpage
\onecolumngrid

\begin{appendix}

\section{Interpolation}
\label{app:Interpolation}

\begin{figure}[h!]
    \centering
    \includegraphics[scale = 1.0]{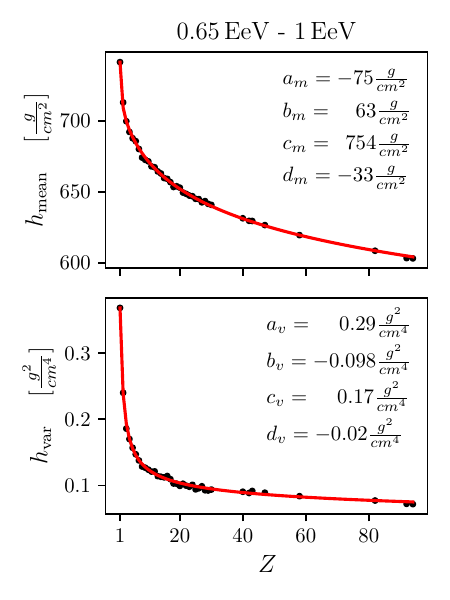}\hspace{1cm}
    \includegraphics[scale = 1.0]{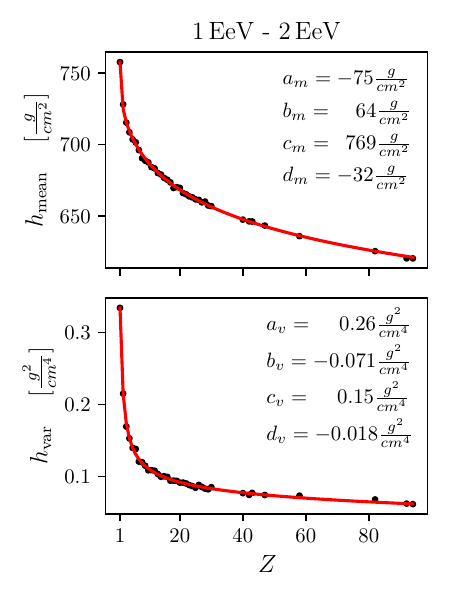}
    \caption{Fits of mean and variance of $z_1$ for $[0.65,1]$ EeV (left) and $[1,2]$ EeV (right) energy bins. The black dots represent simulated values, while the red line indicates the result of the fit.
    }
    \label{fig:z1_interpolation}
\end{figure}

In Fig.~\ref{fig:z1_interpolation} we show the values of the mean (top) and variance (bottom) of the $z_1$ distributions as function of $Z$, for the $[0.65,1]$ (left) and $[1,2]$ (right) energy bins, respectively. The black dots, indicating the values extracted from simulations, can be fitted by two simple function of $Z$, 
\beq\label{eq:interpolation_functions}
\begin{split}
h_{\text{mean}}(x) = \frac{a_m}{x} + \frac{b_m}{x^2} + c_m + d_m \cdot \log(x)\,, \\
h_{\text{var}}(x) = \frac{a_v}{x} + \frac{b_v}{\sqrt{x}} + c_v + d_v \cdot \log(x)\,.
\end{split}
\eeq
As a consequence, we can extract the values of $z_1$, and thus of the distances, for all primaries between $Z=1$ and $Z=94$, without the need to simulate the longitudinal profile of the respective showers. The results of the fit are shown as red lines in Fig.~\ref{fig:z1_interpolation}.

\section{[1,2] bin plots}
\label{app:OtherBin}

\begin{table}[t]
    \centering
    \begin{tabular}{c|c|p{60mm}}
        $d_0$ & $N_{L}$ ($N_{H}$) &  List of atomic numbers $Z$ \\ \hline
        17.7  & 4 (2) &  1,  3, 10, 25, 54, 94    \\ \hline
         7.5  & 8 (5) &  1,  2,  3,  5,  8, 12, 18, 26, 36, 50, 68, 91, 94  \\ \hline
         4.1  & 12 (8)&  1,  2,  3,  4,  5,  7,  9, 11, 14, 17, 21, 26, 31,
        37, 44, 52, 62, 73, 86, 94  \\ \hline
         2.8 & 16 (11) &  1,  2,  3,  4,  5,  6,  7,  8,  9, 11, 13, 15, 17, 20, 23, 26, 30,
       34, 38, 43, 48, 54, 61, 68, 76, 85, 94 \\ \hline
         2.1 & 20 (15) & 1,  2,  3,  4,  5,  6,  7,  8,  9, 10, 11, 12, 13, 14, 16, 18, 20,
       22, 24, 26, 29, 32, 35, 38, 41, 45, 49, 53, 58, 63, 68, 74, 80, 87, 94   \\ \hline
        1.4 & 24 (22) & 1,  2,  3,  4,  5,  6,  7,  8,  9, 10, 11, 12, 13, 14, 15, 16, 17,
       18, 19, 20, 21, 22, 24, 26, 28, 30, 32, 34, 36, 38, 40, 43, 46, 49,
       52, 55, 58, 61, 65, 69, 73, 77, 81, 86, 91
    \end{tabular}
    \caption{Same as Table~\ref{tab:list_primaries_lower_bin1}, for the energy bin $[1,2]$ EeV.}
    \label{tab:list_primaries_bin2}
\end{table}

\begin{figure}[t]
    \centering
    \includegraphics[scale = 0.8]{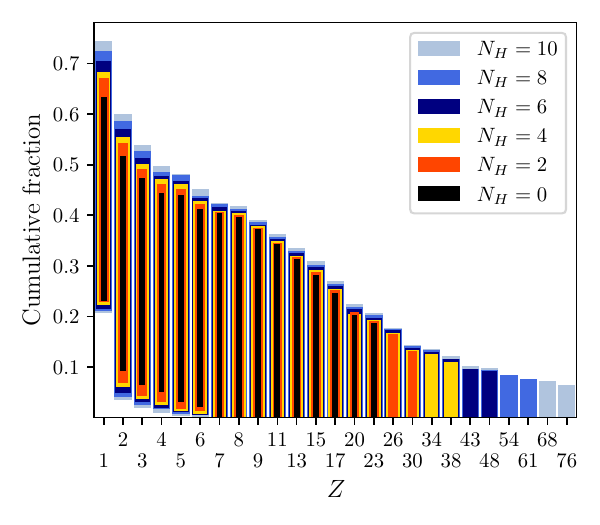}\hspace{1cm}
    \includegraphics[scale = 0.8]{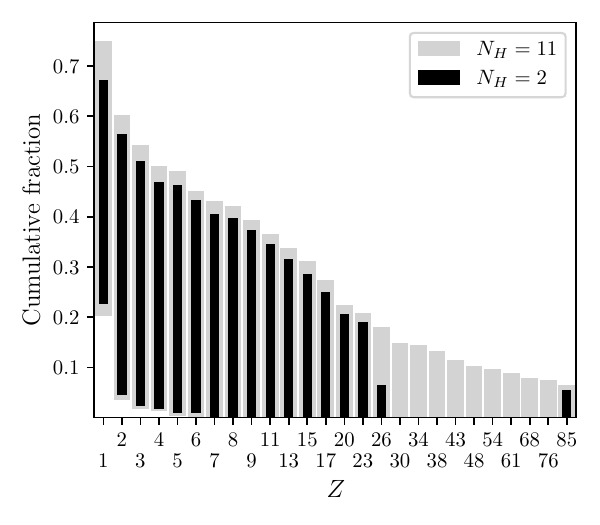}
    \caption{Same as Fig.~\ref{fig:composition_heavy}, for the energy bin $[1,2]$ EeV. }
    \label{fig:composition_heavy_bin2}
\end{figure}

\begin{figure}[t!]
    \centering
    \includegraphics[scale =0.7]{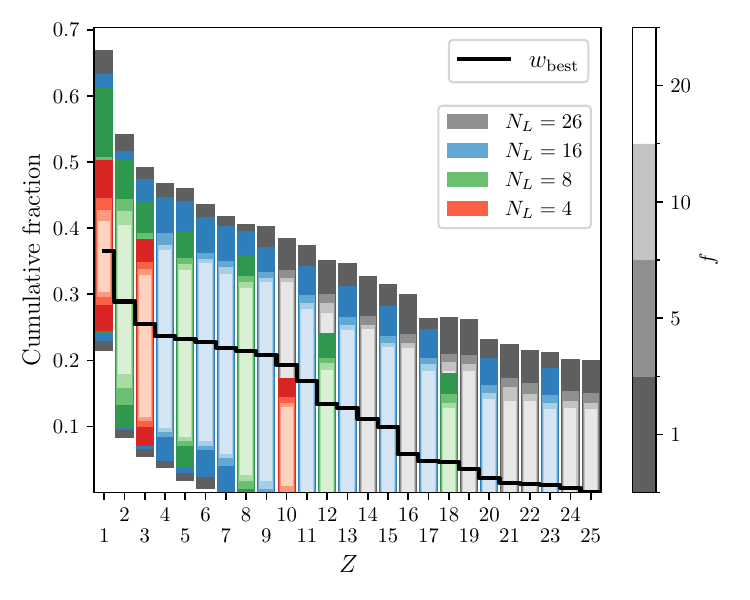}
    \caption{Same as Fig.~\ref{fig:composition:Real:onlyLight}, for the energy bin $[1,2]$ EeV.
    }
    \label{fig:composition:Real:onlyLight2}
\end{figure}

\end{appendix}

\end{document}